\documentclass{ws-procs975x65}

\newcommand{\lwig}{\mbox{\,\raisebox{.3ex}
    {$<$}$\!\!\!\!\!$\raisebox{-.9ex}{$\sim$}\,}}
\newcommand{\gwig}{\mbox{\,\raisebox{.3ex}
    {$>$}$\!\!\!\!\!$\raisebox{-.9ex}{$\sim$}}\,}

\newif\ifhepph

\hepphtrue

\begin{document}

\title{
\ifhepph
\vspace{-3cm}
{\rm\normalsize\rightline{ITP-Budapest 610}\rightline{WUB 04-05}\rightline{DESY 04-014}\rightline{\lowercase{hep-ph/0402102}}}
\vskip 1cm 
\fi
Strong Neutrino-Nucleon Interactions at Ultrahigh Energies 
       as a Solution to the GZK Puzzle\ifhepph\footnote{\uppercase{I}nvited talk given at the 10$^{\rm th}$ 
\uppercase{M}arcel \uppercase{G}ro\ss mann \uppercase{M}eeting, 20-26 \uppercase{J}uly 2003, 
\uppercase{R}io de \uppercase{J}aneiro, \uppercase{B}razil.}\fi}

\author{Z.~Fodor$^{1,2}$, S.~D.~Katz$^{2}$\footnote{\uppercase{O}n leave from \uppercase{I}nstitute
for \uppercase{T}heoretical \uppercase{P}hysics, \uppercase{E}\"otv\"os 
\uppercase{U}niversity, \uppercase{B}udapest, \uppercase{H}ungary.}, 
A.~Ringwald$^3$ and H.~Tu$^3$}

\address{$^1$Institute for Theoretical Physics,\\ E\"otv\"os University,\\ 
H-1518 Budapest, Pf. 32, Hungary}

\address{$^2$Department of Physics, \\
University of Wuppertal, \\
D-42097 Wuppertal, Germany}

\address{$^3$Deutsches Elektronen-Synchrotron DESY, \\ 
D-22603 Hamburg, Germany}  

\maketitle

\abstracts{
After a short review of the ultrahigh energy cosmic ray puzzle  -- 
the apparent observation of cosmic rays originating from cosmological distances 
with energies above the expected Greisen-Zatsepin-Kuzmin cutoff $4\times 10^{19}$~eV --
we consider strongly interacting neutrino scenarios as an especially interesting
solution. We show that all features of the ultrahigh energy cosmic ray spectrum
from $10^{17}$~eV to $10^{21}$~eV 
can be described to originate from a simple power-like injection spectrum of
protons, under the assumption that the neutrino-nucleon cross-section
is significantly enhanced at center of mass energies above $\approx
100$~TeV. In such a scenario, the cosmogenic neutrinos produced during the
propagation of protons through the cosmic microwave 
background initiate air showers in the atmosphere, just
as the protons.
The total air shower spectrum 
induced by protons and neutrinos
shows excellent agreement with the observations.
We shortly discuss TeV-scale 
extensions of the Standard Model which may lead to a 
realization of a strongly interacting neutrino scenario. 
We emphasize, however, that such a scenario may
even be realized within the standard electroweak model: 
electroweak instanton/sphaleron induced processes may get
strong at ultrahigh energies.
Possible tests of strongly interacting neutrino scenarios 
range from observations at cosmic ray facilities and neutrino
telescopes to searches at lepton nucleon scattering experiments.  
}

\section{Introduction}

\begin{figure}
\centerline{\epsfxsize=3.2in\epsfbox{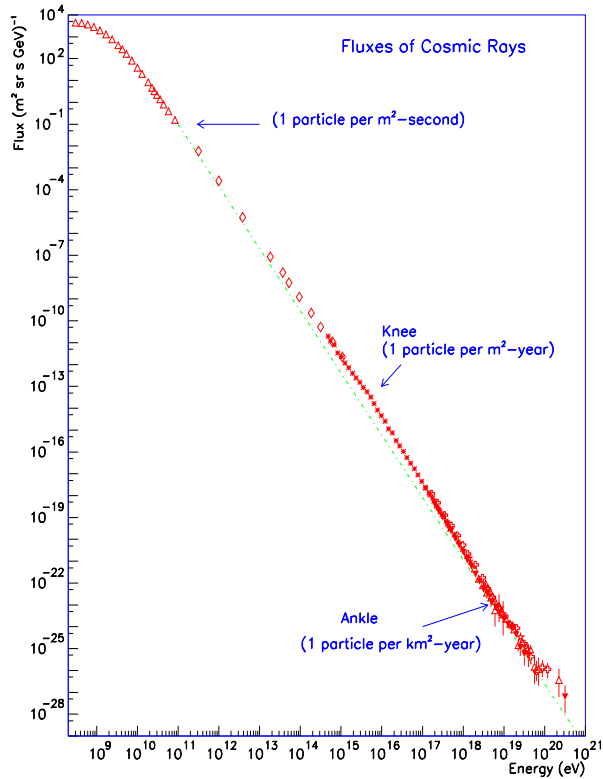}}   
\caption[...]{Compilation of measurements of the flux of 
cosmic rays. The dotted line shows an $E^{-3}$ power-law for comparison. 
Approximate integral fluxes (per steradian) are also shown 
(adapted\cite{Olinto:2000fz} from Ref.\cite{Cronin:vy}).
\label{cr-spectrum}}
\end{figure}

The Earth's atmosphere is continuously bombarded by cosmic particles (``rays''). 
Their measured flux extends over many orders of magnitude in energy  
(cf. Fig.~\ref{cr-spectrum}). At energies above $10^{15}$~eV, they 
are observed in the form of extensive air showers (EAS's), 
initiated by inelastic scattering processes of cosmic particles off atmospheric nucleons.  
Ground-based observatories have measured EAS's with energies up to $E\,\lwig\, 3\times 10^{20}$~eV, 
corresponding to center-of-mass (CM) energies $\sqrt{s}=\sqrt{2 m_p E}\,\lwig\, 750$~TeV, 
where $m_p$ is the proton mass. Therefore, the highest energy cosmic rays 
probe physics beyond the reach of the (Very\cite{vlhc}) Large Hadron Collider\cite{lhc} ((V)LHC), 
with a projected CM energy of $14\,(200)$~TeV.
In this context, it is interesting that the measured cosmic ray flux at 
the highest energies, $E\gwig 10^{20}$~eV, represents a puzzle. 
What is this puzzle about? 

\begin{figure}
\vspace{-0.2cm}
\centerline{\epsfxsize=3.6in\epsfbox{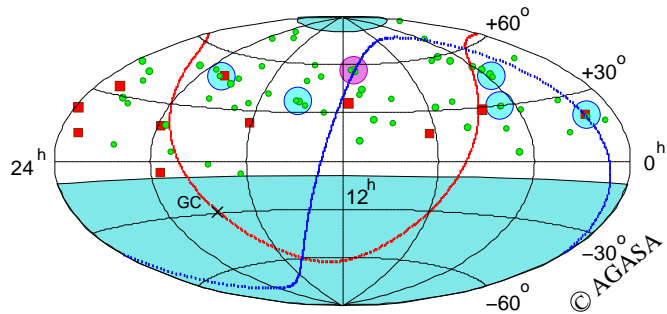}}   
\caption[...]{
Arrival directions of cosmic rays detected by the AGASA and 
Akeno (A20) experiments in equatorial coordinates. Open circles and open squares 
represent cosmic rays with energies $(4 \div 10) \times 10^{19}$~eV, 
and $\geq 10^{20}$~eV, respectively. The galactic and super-galactic planes 
are shown by the red and blue curves, respectively. Large shaded circles indicate event clusters 
within $2.5^{\circ}$. The shaded regions indicate the celestial regions 
excluded by a zenith angle cut of $\leq 45^\circ$. Update\cite{AGASA} (June 24, 2003) of the published 
data from Ref.\cite{Takeda:1999sg}.
\label{arr-dist}}
\end{figure}

It hinges on the circumstantial evidence that 
the cosmic rays above $10^{17.5\div 18.5}$~eV originate 
from cosmological distances (for a recent review, see Ref.\cite{Anchordoqui:2002hs}). 
This evidence is largely based on 
the apparent large-scale isotropy in the arrival directions of cosmic rays 
(cf. Fig.~\ref{arr-dist}). 
Moreover, whereas there are only very few -- if any -- nearby source candidates, 
plausible astrophysical sources are most likely to be found only at 
cosmological distances. 

\begin{figure}
\vspace{-0.9cm}
\centerline{\epsfxsize=3.8in\epsfbox{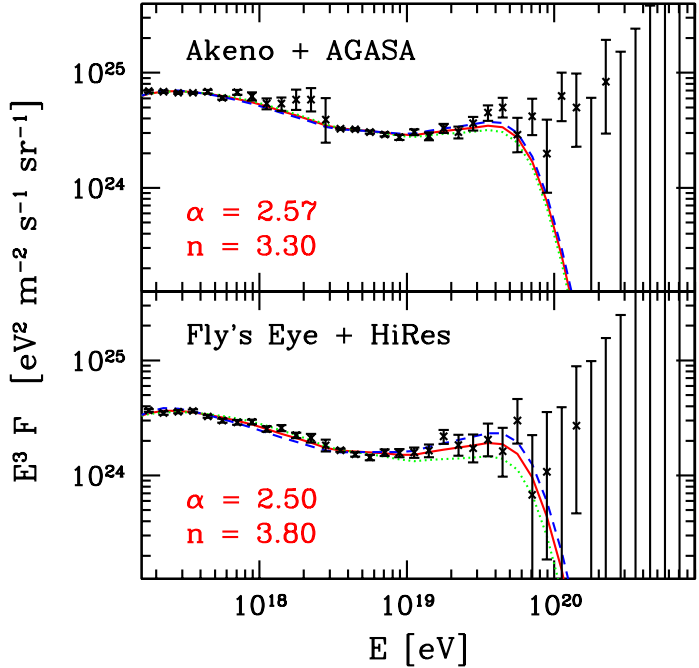}}   
\caption[...]{
Ultrahigh energy cosmic ray data with their statistical errors 
(top: combination of Akeno\cite{Nagano:1991jz} and AGASA\cite{Takeda:1998ps} data; 
bottom: combination of Fly's Eye\cite{Bird:yi} and 
HiRes\cite{Abu-Zayyad:2002ta} data) and the predictions arising from 
a power-law emissivity distribution~(\ref{source-emissivity}) corresponding to sources which
are uniformly distributed at cosmological distances.    
The best fits between $E_-=10^{17.2}$~eV and $E_+=10^{20}$~eV 
are given by the solid lines and correspond
to the indicated values of the parameters $\alpha$ and $n$ in the source emissivity
distribution.  
The 2-sigma variations corresponding to the minimal (dotted) and 
maximal (dashed) fluxes are also shown.  
Other parameters of the analysis were
$E_{\rm max} = 3 \times 10^{21}$~eV, $z_{\rm min} = 0.012$, and $z_{\rm max}=2$. 
From Ref.\cite{Fodor:2003ph}.
\label{uhecr-data}}
\end{figure}

If the highest energy cosmic rays are nucleons (or nuclei), 
if their sources are indeed uniformly distributed at cosmological
distances, and if their injection spectra are power-laws in energy  
-- a reasonable assumption, in view of the measured spectrum in Fig.~\ref{cr-spectrum} which appears to be 
approximately of (broken) power-law type over many order of magnitude in energy --  
then their total flux arriving at Earth should show a pronounced drop 
above the Greisen-Zatsepin-Kuzmin\cite{Greisen:1966jv} (GZK) ``cutoff'' $E_{\rm GZK}=4\times 10^{19}$~eV. 
This is due to the fact that, 
above this energy, the universe becomes opaque to high energy nucleons (and nuclei), due to 
inelastic hadronic scattering processes with the cosmic microwave background (CMB) 
photons. The GZK cutoff is, however, not seen in the data, at least not in 
a significant manner (cf. Fig.~\ref{uhecr-data}). Correspondingly, the events above 
$10^{20}$~eV in Fig.~\ref{uhecr-data} should originate from small distances 
below $50$~Mpc, the typical interaction length of nucleons above $E_{\rm GZK}$. 
However, no source within a distance of $50$~Mpc is known in the arrival directions of
the post-GZK events\footnote{The dominant radio galaxy M87 in the Virgo cluster, at 
a distance of about $20$~Mpc,  
has been a source candidate for a long time\cite{Ginzburg:63}. 
The major difficulty with this idea is the isotropy of the arrival distribution. 
It might be overcome by invoking a particular galactic magnetic field originating 
from a ``galactic wind''\cite{Ahn:1999jd}. Criticisms of this model\cite{Billoir:2000wi} have been 
addressed in Ref.\cite{Biermann}.}. 
The basic puzzle is: 
if the sources of ultrahigh energy cosmic rays are indeed at cosmological distances, 
how could they reach us with energies above $10^{20}$~eV?

At the relevant energies, among the known particles only neutrinos can 
propagate without 
significant energy loss from cosmological distances to us. 
It is this fact which led, on the one hand, 
to scenarios invoking 
hypothetical -- beyond the Standard Model --  
strong interactions of ultrahigh energy 
cosmic 
neutrinos\cite{Beresinsky:qj} and, on the other hand, to the Z-burst 
scenario\cite{Weiler:1997sh,Fodor:2001qy}.

In the latter, ultrahigh energy cosmic neutrinos (UHEC$\nu$'s) 
produce Z-bosons through annihilation with the relic neutrino background from the big bang.
On Earth, we observe the air showers initiated by the protons and photons 
from the hadronic decays of these Z-bosons. Though the required ultrahigh energy
cosmic neutrino flux\cite{Fodor:2001qy} is smaller than present upper 
bounds\cite{Kravchenko:2003tc}, it is not easy to 
conceive a production mechanism yielding a sufficiently large one. 
In the near future, UHEC$\nu$ detectors, such as 
the Pierre Auger Observatory\cite{Auger}, IceCube\cite{IceCube}, ANITA\cite{ANITA}, 
EUSO\cite{EUSO}, OWL\cite{OWL}, and SalSA\cite{Gorham:2001wr} 
can directly confirm or exclude this scenario\cite{Eberle:2004ua}.

Scenarios based on strongly interacting neutrinos, on the other hand, 
are based on the observation that the flux of 
neutrinos originating from the decay of the pions produced during the
propagation of nucleons through the CMB\cite{Beresinsky:qj,Stecker:1979ah,Yoshida:pt,Protheroe:1995ft} 
-- the cosmogenic neutrinos -- shows a nice agreement with the observed ultrahigh energy cosmic ray (UHECR) 
flux above 
$E_{{\rm GZK}}$.
Assuming a large enough neutrino-nucleon cross-section
at these high energies, these neutrinos could initiate extensive
air showers high up in the atmosphere, like hadrons, and
explain the existence of the post-GZK events\cite{Beresinsky:qj}.
This large cross-section
is usually ensured by new types of TeV-scale 
interactions beyond the Standard Model,
such as 
arising through gluonic bound state 
leptons\cite{Bordes:1997bt}, through 
TeV-scale grand unification with leptoquarks\cite{Domokos:2000dp}, 
through Kaluza-Klein modes from compactified extra 
dimensions\cite{Domokos:1998ry} 
(see, however, Ref.\cite{Kachelriess:2000cb}), 
or through $p$-brane production in models with warped extra dimensions\cite{Ahn:2002mj}
(see, however, Ref.\cite{Anchordoqui:2002it});  
for earlier and further proposals, see Ref.\cite{Domokos:1986qy} and 
Ref.\cite{Barshay:2001eq}, respectively. 

In this review, we discuss strongly interacting
neutrino scenarios as a possible solution to the GZK puzzle. 
We present  a detailed statistical analysis of the agreement between observations and 
predictions from such scenarios\cite{Fodor:2003bn}.
Moreover, we 
emphasize an example which -- in contrast to previous proposals -- 
is based entirely on the Standard Model of particle physics. 
It exploits non-perturbative electroweak instanton-induced 
processes
for the interaction of cosmogenic neutrinos with nucleons in the atmosphere, 
which may have a sizeable cross-section above a threshold energy 
$E_{\rm th}={\mathcal O}( (4\pi m_W/\alpha_W )^2)/(2 m_p) = {\mathcal O}( 10^{18})$~eV, 
where 
$m_W$ denotes the W-boson mass and $\alpha_W$ the electroweak fine structure 
constant\cite{Aoyama:1986ej,Morris:1991bb,Ringwald:2002sw}.

Our scenario is based on a standard power-like primary spectrum of
protons injected from sources at cosmological distances.
After propagation through the CMB, these protons will have energies below $E_{{\rm GZK}}$, so
they can well describe the low energy part of the UHECR spectrum.
The cosmogenic neutrinos interact with the atmosphere and thus
give a second component to the UHECR flux, which describes
the high energy part of the spectrum. The relative normalization
of the proton and neutrino fluxes is fixed in this scenario, so
the low and high energy parts of the spectrum are explained 
simultaneously without
any extra normalization.
Details of this analysis can be found in Ref.\cite{Fodor:2003bn}.   

The structure of this review is as follows. In the next section, 
we review our procedure to infer the fluxes of protons and cosmogenic neutrinos at 
Earth, from an assumed injection spectrum at the sources. In Sect.~\ref{inst-spect}, 
various possibilities, including the one exploiting electroweak instantons, 
for a large neutrino-nucleon cross-section at high energies are discussed, and the induced
air shower rate is calculated. In Sect.~\ref{comparison}, 
we present a comparison of the predictions with the observations and a 
determination of the goodness of the fit. Possible further tests are mentioned in 
Sect.~\ref{tests},  while conclusions are given in Sect.~\ref{conclusions}.

\section{\label{fluxes}Proton and cosmogenic neutrino fluxes}

Our analysis\cite{Fodor:2003bn} is based on the assumption of a 
power-law emissivity distribution corresponding to uniformly distributed
sources. The emissivity  is defined as the number of 
protons per co-moving volume per unit of time and per unit of energy, injected
into the CMB 
with energy $E_i$ and  characterized by a 
spectral index $\alpha$  and a redshift ($z$) evolution index $n$,
\begin{equation}
\label{source-emissivity}
{\mathcal L}_p =j_0\,E^{-\alpha}_i\,\left(1+z\right)^n\,\theta(E_{\rm max}-E_i)\,
\theta(z-z_{\rm min})\,\theta(z_{\rm max}-z)\,.
\end{equation}
Here, $j_0$ is a normalization factor, which will be fixed by the observed flux. 
The parameters $E_{\rm max}$ and  $z_{\rm min/max}$ 
have been introduced to take into account certain possibilities such as the existence 
of a maximal energy, which can be reached through astrophysical 
accelerating processes in a bottom-up scenario, and the absence of nearby/very early sources,
respectively. 
Our predictions are quite insensitive to the
specific choice for 
$E_{\rm max}$, $z_{\rm min}$, and $z_{\rm max}$, within their anticipated
values. 
The main sensitivity arises from the spectral parameters $\alpha$ and $n$, 
for which we determine the 1- and 2-sigma confidence regions in 
Sect.~\ref{comparison}.

\begin{figure}[b]
\vspace{-0.9cm} 
\centerline{\epsfxsize=3.5in\epsfbox{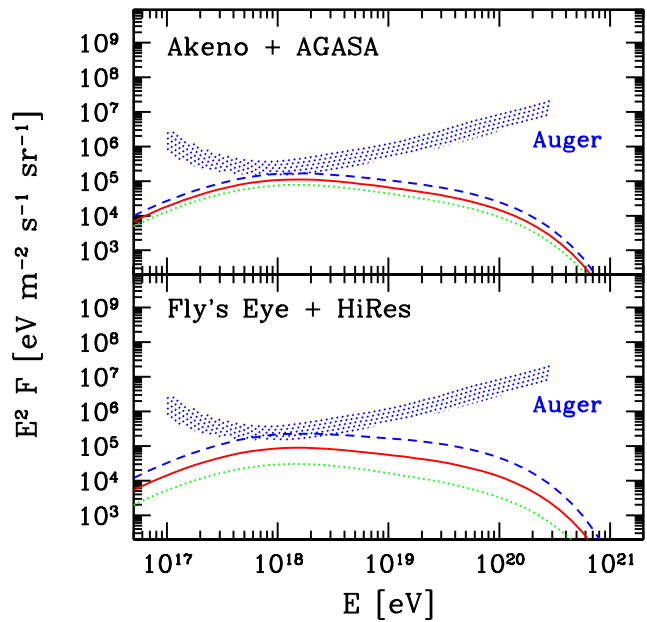}}   
\caption[...]{
Predicted cosmogenic neutrino fluxes per flavor, $F_{\nu_\ell}+F_{\bar\nu_\ell}$, $\ell =e,\mu,\tau$,
originating from a power-law proton source emissivity distribution~(\ref{source-emissivity}) and 
corresponding to the predicted UHECR fluxes in Fig.~\ref{uhecr-data}.   
The ``best'' predictions for the neutrino spectra   
are given by the solid lines. 
The 2-sigma variations corresponding to the minimal (dotted) and 
maximal (dashed) fluxes are also shown. 
The dotted band labelled by Auger represents the expected 
sensitivity of the Pierre Auger Observatory to $\nu_\tau +\bar\nu_\tau$, 
corresponding to one event
per year per energy decade\cite{Bertou:2001vm}.
From Ref.\cite{Fodor:2003ph}. 
\label{cosmogenic}}
\end{figure}

The propagation of particles can be 
described\cite{Yoshida:pt,Bahcall:1999ap,Fodor:2000yi}
by $P_{b|a} (z,E_i;E )$ functions, which give the expected number of 
particles of type $b$ 
above the threshold energy $E$ if one particle of type $a$ 
started at  a redshift ``distance'' $z$ with energy $E_i$. 
With the help of these propagation functions, 
the differential flux of protons ($b=p$) 
and cosmogenic neutrinos 
($b=\nu_i, \bar\nu_i$) at Earth
can be written as   
\begin{equation}
\label{flux-earth}
F_{b} ( E ) =\frac{1}{4\pi}
 \int_0^\infty {\rm d}E_i \int_0^\infty \frac{{\rm d}z}{H(z)} 
\,\frac{-\,\partial P_{b|p}(z,E_i;E)}{\partial E}
\,\frac{{\mathcal L}_p (r,E_i)}{1+z}\,.
\end{equation}
In our analysis, we took     
$z_{\rm max} = 2$ (cf.~Ref.~\cite{Waxman:1995dg}), while 
we choose $z_{\rm min}=0.012$ in order to take into account the fact that within 
$50$~Mpc there are apparently no astrophysical sources of UHECR's. 
We used the expression 
$H^2(z) = H_0^2\,\left[ \Omega_{M}\,(1+z)^3 
+ \Omega_{\Lambda}\right]$ 
for the relation of the Hubble expansion rate at redshift $z$ to the present one.
Uncertainties of the latter, $H_0=h$ 100 km/s/Mpc, with\cite{Hagiwara:fs} 
$h=(0.71\pm 0.07)\times^{1.15}_{0.95}$, 
are included. 
$\Omega_{M}$ and $\Omega_{\Lambda}$, with $\Omega_M+\Omega_\Lambda =1$, are the present 
matter and vacuum energy densities in terms of the critical density. As default values we choose
$\Omega_M = 0.3$ and $\Omega_\Lambda = 0.7$, as favored today. Our results
turn out to be rather 
insensitive to the precise values of the cosmological parameters.

We calculated $P_{b|a}(z,E_i;E)$ in two steps. 
{\em i)} First, the SOPHIA
Monte-Carlo program\cite{Mucke:1999yb} was 
used for the simulation of photohadronic processes of protons with the CMB photons. 
For $e^+e^-$ pair production, we used the continuous energy loss approximation, since the
inelasticity is very small ($\approx 10^{-3}$).
We calculated
the $P_{b|a}$ functions for ``infinitesimal'' steps as a function of 
the redshift $z$.
{\em ii)} 
We multiplied the corresponding infinitesimal probabilities  
starting at a redshift $z$ down to Earth with $z=0$. 
The details of the calculation of the $P_{b|a}(z,E_i;E )$ functions for
protons, neutrinos, charged leptons, and photons will be published
elsewhere\cite{P:xxx}.

Since the propagation functions are of universal usage, we decided
to make the latest versions of $-\partial P_{b|a}/\partial E$ 
available for the public via the World-Wide-Web URL 
www.desy.de/\~{}uhecr \,.

As an illustration of the outcome of our propagation codes, we display in 
Fig.~\ref{uhecr-data} the predictions for the proton flux at Earth,  
originating from a power-like source emissivity distribution~(\ref{source-emissivity}) 
with particular $\alpha, n, \ldots$ values indicated on the figure and in its caption. 
A nice fit of the data can be obtained apparently for energies below 
$\lwig\, 4\times 10^{19}\ {\rm eV} =E_{\rm GZK}$ -- more on this in Sect.~\ref{comparison}.  
The associated predicted cosmogenic neutrino flux, for the same parameter values, 
is displayed in Fig.~\ref{cosmogenic}.   

\section{\label{inst-spect}Spectrum of neutrino-induced air showers}

The main assumption of strongly interacting neutrino scenarios is that the neutrino-nucleon
cross-section $\sigma_{\nu N}^{\rm tot}$ suddenly becomes much larger than $\approx 1$~mb above 
center of mass energies $\sqrt{s} \approx 100$~TeV.
In this case, the corresponding neutrino interaction length 
$\lambda_\nu \equiv m_p/\sigma_{\nu N}^{\rm tot}$, with 
$\sigma_{\nu N}^{\rm tot}=\sigma_{\nu N}^{\rm cc}+ \sigma_{\nu N}^{\rm new}$,
falls below $X_0=1031$~g/cm$^2$ -- the vertical depth 
of the atmosphere at sea level -- above the neutrino threshold 
energy $\approx 10^{19}$~eV.  
Here $\sigma_{\nu N}^{\rm cc}$ and
$\sigma_{\nu N}^{\rm new}$ denote the charged current and the new contribution
to the cross-section.
Above the neutrino threshold energy, the atmosphere becomes opaque to cosmogenic neutrinos 
and most of them will end up as 
air showers. Quantitatively, this fact can be described by   
\begin{equation}
\label{shower-rate}
F'_\nu (E) =
F_\nu (E)\,
\left[ 1 - {\rm e}^{- \frac{X(\theta )}{\lambda_\nu (E)}}\right]
= F_\nu (E)\times \left\{ 
\begin{array}{ccl}
\frac{X(\theta )}{\lambda_\nu (E)}\  & \ {\rm for}\  & \lambda_\nu (E)\gg X(\theta )\\[1.5ex]
1 \  & \ {\rm for}\  & \lambda_\nu (E)\ll X(\theta )
\end{array}
\right. 
\,,
\end{equation}
which gives 
the spectrum 
of neutrino-initiated air showers, 
for an incident cosmogenic neutrino flux  
$F_\nu =\sum_i [F_{\nu_i}+F_{\bar\nu_i}]$ 
from Eq.~(\ref{flux-earth}), 
in terms of the atmospheric depth $X(\theta )$,  
with $\theta$ being the zenith angle. 

\begin{figure}
\centerline{\epsfxsize=3.3in\epsfbox{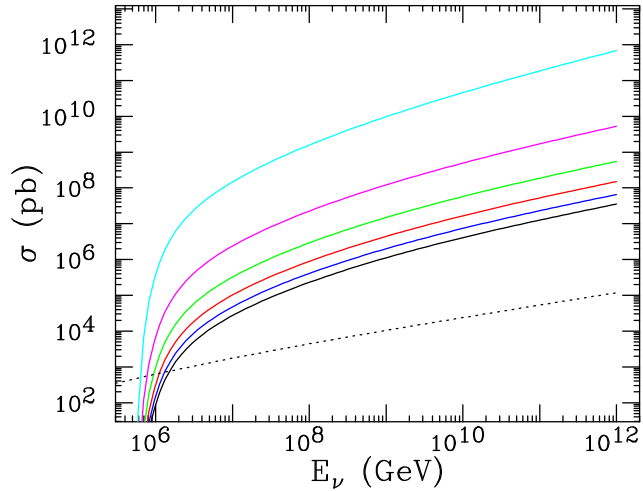}}   
\caption[...]{
Total cross section $\sigma (\nu N \to p{\rm -brane})$ in a model with $n=6$ 
extra dimensions, out of which $m$ have a size $L/L_{*} = 0.25$ in terms of the fundamental Planck length 
$L_\ast=M_\ast^{-1}$, for  $m = 0, \ldots, n-1$ from below. The fundamental Planck scale 
$M_D=[(2\pi)^n/8\pi]^{1/(n+2)} M_\ast$ has been chosen as $M_D = M_p^{\rm min} =1$~TeV in terms
of the minimum $p$-brane mass $M_p^{\rm min}$. 
The Standard Model charged current cross-section
$\sigma(\nu N \to \ell X)$ is also shown (dotted). From Ref.\cite{Anchordoqui:2002it}.
\label{brane-cs}}
\end{figure}

Such suddenly increasing cross-sections have been proposed
in various models involving physics beyond the Standard 
Model\cite{Bordes:1997bt,Domokos:2000dp,Domokos:1998ry,Kachelriess:2000cb,Ahn:2002mj,Anchordoqui:2002it,Domokos:1986qy,Barshay:2001eq}.
Among the usual suspects are TeV-scale gravity scenarios with large or warped extra dimensions\cite{Antoniadis:1990ew}. 
In those, the neutrino-nucleon cross-section may be greatly enhanced compared to the Standard Model 
one. As an example, we demonstrate in Fig.~\ref{brane-cs} that $p$-brane production in neutrino-nucleon 
scattering\cite{Ahn:2002mj,Anchordoqui:2002it} may reach a cross-section of $\approx 10$~mb at $\approx 10^{19}$~eV,  
depending on the parameters of the model. This is in contrast to 
microscopic black hole ($\equiv 0$-brane) production\cite{Giddings:2001bu} which has generically too small 
a cross-section\cite{Feng:2001ib} to solve the GZK puzzle, within the allowed parameter ranges. 

\begin{figure}[b]
\vspace{-2cm}
\centerline{\epsfxsize=3.9in\epsfbox{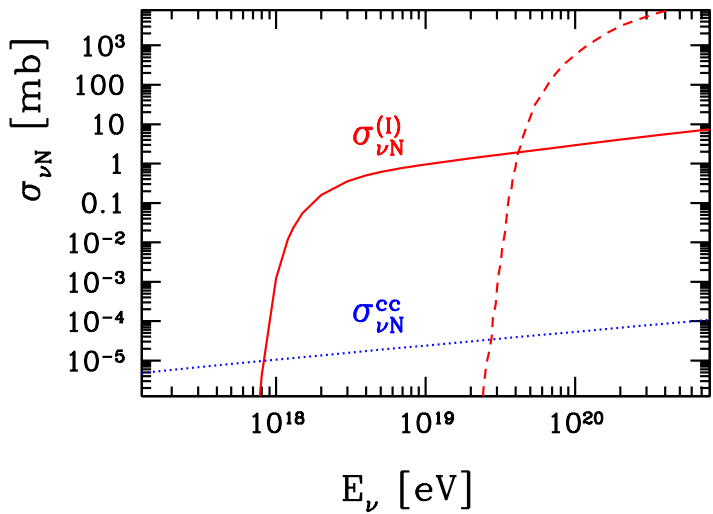}}
\vspace{-1.2cm}  
\caption[...]{
Predictions of the electroweak instanton-induced neutrino-nucleon cross-section 
$\sigma_{\nu N}^{(I)}$ (solid\cite{Fodor:2003bn} and dashed\cite{Han:2003ru}) 
in comparison with the charged current cross-section 
$\sigma_{\nu N}^{\rm cc}$ (dotted) from Ref.\cite{Gandhi:1998ri}, as a function of 
the neutrino energy $E_\nu$ in the nucleon's rest frame. 
\label{cross-nuN}}
\end{figure}

In Fig.~\ref{cross-nuN}, we show another example for a strong enhancement in the 
neutrino-nucleon cross-section, which is based entirely 
on the Standard Model, exploiting non-perturbative electroweak 
instanton-induced processes\cite{Aoyama:1986ej,Morris:1991bb,Ringwald:2002sw}.
According to the estimates presented in Fig.~\ref{cross-nuN}, the electroweak instanton-induced 
neutrino-nucleon cross-section appears to have a threshold-like 
behavior at $E_{\rm th}={\mathcal O}( (4\pi m_W/\alpha_W )^2)/(2 m_p) = {\mathcal O}( 10^{18})$~eV, 
above which it quickly rises above 1~mb.  
Our quantitative analysis in Ref.\cite{Fodor:2003bn} was based on 
the cross-section from Ref.\cite{Ringwald:2002sw} (solid line in Fig.~\ref{cross-nuN}), 
however it is quite insensitive to the exact form of it as long as it rises abruptly far above 1~mb. 
Note that such a behaviour is consistent with present upper bounds
on electroweak instanton-induced cross-sections\cite{Bezrukov:2003er,Ringwald:2003ns}.
However, it is fair to say that there are substantial uncertainties in the 
predictions in Fig.~\ref{cross-nuN}: the absolute size of the cross-section above
the threshold energy may well be unobservably small.

\section{\label{comparison}Comparison with UHECR data}

The predicted air shower rate induced by protons and neutrinos
is given by
\begin{equation}
\label{flux-pred}
F_{\rm pred} (E; \alpha , n, E_{\rm max},  z_{\rm min}, z_{\rm max}, j_0 ) 
= F_{p} ( E; \ldots  ) + F'_{\nu} (E; \ldots )\,.
\end{equation}
In Ref.\cite{Fodor:2003bn}, we performed a statistical 
analysis to compare the prediction (\ref{flux-pred}), within the electroweak 
instanton scenario from Fig.~\ref{cross-nuN} (solid),  
with the observations and presented a measure for the goodness of the scenario. 
We gave the best fit to the observations and the 1- and 2-sigma
confidence regions in the ($\alpha$,$n$) plane.

\begin{figure}
\vspace{-0.5cm}
\centerline{\epsfxsize=3.9in\epsfbox{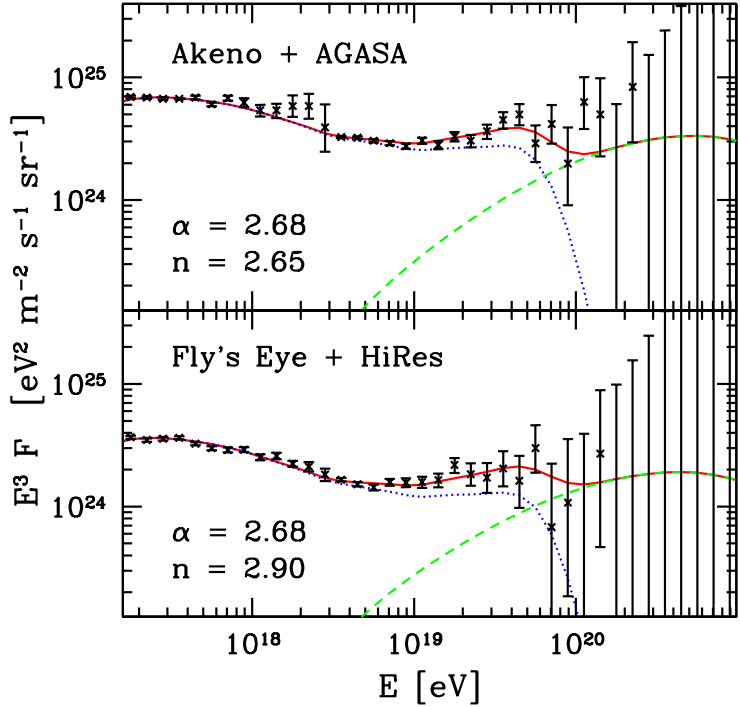}}
\caption[...]{
Ultrahigh energy cosmic ray data 
(Akeno + AGASA on the upper panel and Fly's Eye + HiRes on the lower
panel) and their best fits (solid) within the electroweak instanton scenario 
from Fig.~\ref{cross-nuN} (solid)   
for $E_{\rm max}=3\times 10^{22}$~eV, $z_{\rm min}=0.012$, $z_{\rm max}=2$, 
consisting of a proton component (dotted) plus a cosmogenic 
neutrino-initiated component (dashed). 
From Ref.\cite{Fodor:2003bn}.  
\label{fit}}
\end{figure}

To start the analysis, we had to convert the measured fluxes, which    
UHECR collaborations usually publish in a binned form, into event numbers in each bin. 
We used the most recent results of the HiRes and AGASA
collaborations and did our analysis separately with both data sets.
We concentrated on the energy range $10^{17.2}$~eV -- $10^{21}$~eV which is divided
into 38 equal logarithmic bins. 
In the low energy region, there are no 
published results available from AGASA and only low statistics results
from HiRes-2.
Therefore, we included the results of the predecessor 
collaborations -- Akeno\cite{Nagano:1991jz} and Fly's Eye, 
respectively -- into the analysis. 
With a small normalization correction,  
it was possible to continuously 
connect the AGASA data\cite{Takeda:1998ps} with the Akeno ones and the 
HiRes-1 monocular data\cite{Abu-Zayyad:2002ta} with 
the Fly's Eye stereo ones\cite{Bird:yi},
respectively (cf. Figs.~\ref{uhecr-data} and \ref{fit}).  

The goodness of the scenario was 
determined by a statistical analysis. 
We determined the compatibility of different
($\alpha$,$n$) pairs (cf. Eqs.~(\ref{source-emissivity}) and (\ref{flux-pred})) with the experimental data.
For some fixed ($\alpha$,$n$) pair, the expected number of events in 
individual bins are (${\bf \lambda}=\{\lambda_1,...,\lambda_r\}$ with
$r$ being the total number of bins (in our case 38).
The probability of getting an experimental outcome 
${\bf k}=\{k_1,...k_r\}$ (where $k_i$ are non-negative integer numbers)
is given by the probability distribution function $P({\bf k})$,
which is just the product of Poisson
distributions for the individual bins.
We also included the $\approx 30\%$ overall energy uncertainty into the $P({\bf k})$ 
probability distribution. 
We denote the experimental result 
by ${\bf s}=\{s_1,...s_r\}$, where the $s_i$-s are
non-negative, integer numbers. 
The ($\alpha$,$n$) pair is compatible with the 
experimental results if 
\begin{equation}\label{summation}
\sum_{{\bf k}|P({\bf k})>P({\bf s})}P({\bf k})<c\,.
\end{equation}
For a 1-(or 2-)sigma compatibility one takes c=0.68 
(or c=0.95), respectively. 
The best fit is found by minimizing the sum on the left hand side.

\begin{figure}
\centerline{\epsfxsize=2.8in\epsfbox{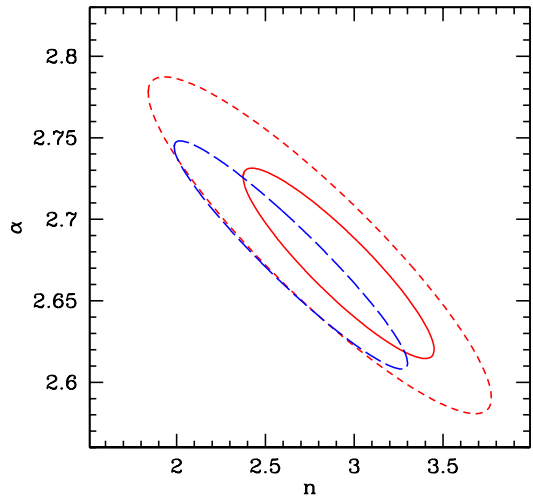}}   
\vspace{-0.5cm}
\caption[...]{
Confidence regions in the $\alpha$--$n$ plane 
for fits to the Akeno + AGASA data (2-sigma (long dashed))
and to the Fly's Eye + HiRes 
data (1-sigma (solid); 2-sigma (short-dashed)), respectively, within 
the electroweak instanton scenario from Fig.~\ref{cross-nuN} (solid), 
for $E_{\rm max}\gwig 3\times 10^{21}$~eV, $z_{\rm min}\geq 0$, 
$z_{\rm max}=2$.  
From Ref.\cite{Fodor:2003bn}. 
\label{confidence}}
\end{figure}

Figure~\ref{fit} shows our best fits for the AGASA and for 
the HiRes UHECR data. The best fit values are $\alpha=2.68(2.68)$ and 
$n=2.65(2.9)$, for AGASA(HiRes), within the electroweak instanton scenario
from Fig.~\ref{cross-nuN} (solid). 
We can see very nice agreement with the data
within an energy range of nearly four orders of magnitude. 
The fits are insensitive to the value of $E_{\rm max}$ as far as
we choose a value above $\approx 3\times 10^{21}$~eV.
The shape of the curve between $10^{17}$~eV and $10^{19}$~eV is mainly
determined by the redshift evolution index $n$. At these energies the
universe is already transparent for protons created at $z\approx 0$, while
protons from sources with larger redshift accumulate in this region.
The more particles are created at large distances -- i.e. the larger $n$ is -- 
the stronger this accumulation should be. In this context, we note that 
the data seem to confirm our implicit assumption that the extragalactic uniform UHECR 
component begins to dominate over the galactic one already at $\approx 10^{17}$~eV. 
If we, alternatively, start our fit only at $10^{18.5}$~eV -- corresponding to the 
assumption that the galactic component dominates up to this energy -- 
we find, however, also a very good fit, with a very mild dependence on $n$ and the same best
fit values for $\alpha$, with a bit larger uncertainties.
The peak around $4\times 10^{19}$~eV in Fig.~\ref{fit} shows the accumulation of particles
due to the GZK effect. Neutrinos start to dominate over protons at around 
$10^{20}$~eV.

It is important to note that, if we omit the neutrino component, 
then the model is ruled out on the 3-sigma level for both experiments. 
This is due to the fact that we excluded nearby sources by setting
$z_{\rm min} \neq 0$ (see also Ref.\cite{Kachelriess:2003yy}). 
The choice $z_{\min}=0$ makes the HiRes data 
compatible with a proton-only scenario on the 2-sigma level 
(see also Refs.\cite{Abu-Zayyad:2002ta,Bahcall:2002wi}). 

Figure~\ref{confidence} 
displays the confidence regions in the ($\alpha$,$n$)
plane
for AGASA and HiRes. The scenario is consistent on the 2-sigma level with
both experiments. For HiRes, the compatibility is even true on the 1-sigma level.
It is important to note that both experiments favor the same values for $\alpha$ and 
$n$, demonstrating their mutual compatibility on the 2-sigma level 
(see also Ref.\cite{DeMarco:2003ig}). 
If we ignore the energy uncertainty in the determination of the goodness 
of the fit, they turn out to be inconsistent.

\begin{figure}
\centerline{\epsfxsize=3.5in\epsfbox{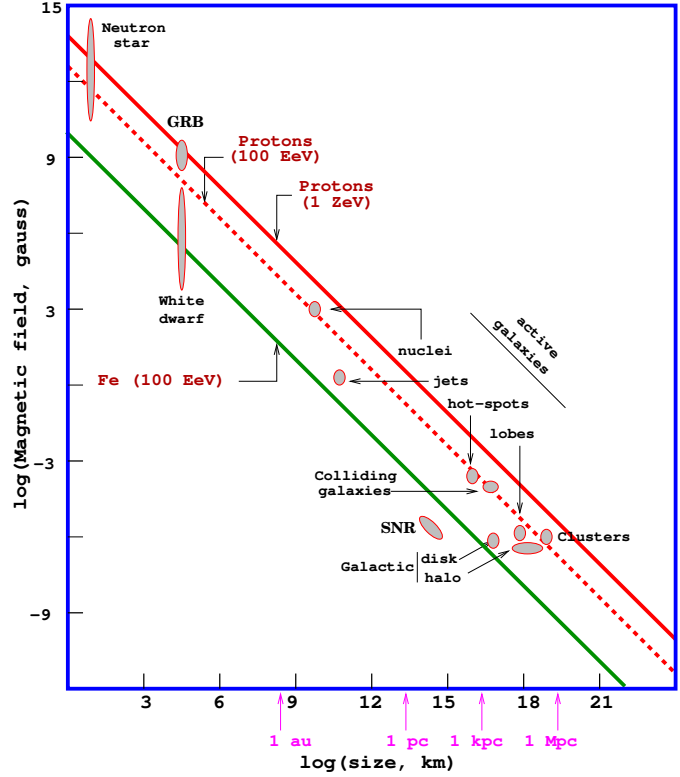}}   
\vspace{-1.cm}
\caption[...]{
The Hillas diagram showing size and magnetic field strengths of 
possible sites of particle acceleration. Objects below the diagonal lines 
(from top to bottom), derived from the Hillas criterion\cite{Hillas:1984} 
$E_{\rm max} \sim 2 Ze\,B\, r$ for the maximum energy acquired by
a particle of charge $Ze$ traveling in a medium of size $r$ with a magnetic field $B$,  
cannot shock accelerate protons above $ 10^{21}$~eV, above $ 10^{20}$~eV and iron nuclei above 
$10^{20}$~eV, respectively.
(This version of the picture is courtesy of Murat 
Boratav).
\label{hillas-plot}}
\end{figure}

Finally, let us emphasize that the same fit results are valid
for all strongly interacting neutrino scenarios, as long as the neutrino-nucleon
cross-section has a similar threshold-like behavior as in 
Figs.~\ref{brane-cs} and \ref{cross-nuN}, with a neutrino threshold energy 
$\lwig\, 4\times 10^{19}$~eV and a cross-section $\gwig\, 1$~mb above 
threshold. It is also important to note that the energy requirements on 
the sources of the primary protons are comparatively mild. To obtain a 
good fit, we need $E_{\rm max}\gwig\, 3\times 10^{21}$~eV. An inspection of 
the Hillas diagram in Fig.~\ref{hillas-plot} reveals that there are a number
of reasonable astrophysical source candidates, notably  neutron stars 
and gamma ray bursters (GRB's), which may provide the necessary conditions to 
accelerate protons to the required energies by conventional
shock acceleration.

\section{\label{tests}Further tests}

There are a number of further possible tests of strongly interacting neutrino scenarios, 
ranging from astroparticle tests, which include searches at EAS arrays and neutrinos telescopes,
to laboratory tests at present and future accelerators. 
We will review some of those in this Section. 

\subsection{Astroparticle tests}

\subsubsection{Searches at EAS arrays}

One possibility to test the ultrahigh energy neutrino component in the EAS data is to study the 
zenith angle dependence of the events in the $10^{18\div 20}$~eV range, which
will reflect the energy dependence of the neutrino-nucleon cross-section\cite{Morris:1991bb,Berezinsky:kz}.  
Near the threshold energy in strongly interacting neutrino scenarios, there
will be always a range of energies where the cross-section is already sizable,
but does not yet reach hadronic values (cf. Figs.~\ref{brane-cs} and \ref{cross-nuN}), 
in particular, where $\sigma_{\nu N}^{\rm tot}\,\lwig\, 0.56$~mb, corresponding to the atmospheric
depth at larger zenith angles, $\theta\,\gwig\, 70^{\circ}$. 
Therefore, for these energies, neutrino-initiated showers can be searched for at cosmic ray facilities 
by looking for quasi-horizontal air showers\cite{Morris:1991bb}, $\theta\, \gwig\, 70^{\circ}$. 
We have checked in Ref.\cite{Fodor:2003bn} that the rate from our electroweak instanton prediction 
(cf. Fig.~\ref{cross-nuN}) is consistent with observational constraints found by the Fly's Eye\cite{Baltrusaitis:mt}
and AGASA\cite{Yoshida:2001icrc} collaborations. For the case of $p$-brane production, 
such constraints can be avoided in warped extra dimension scenarios with fine-tuned 
sizes\cite{Anchordoqui:2002it}.  

\begin{figure}
\vspace{-0.5cm}
\centerline{\epsfxsize=3.1in\epsfbox{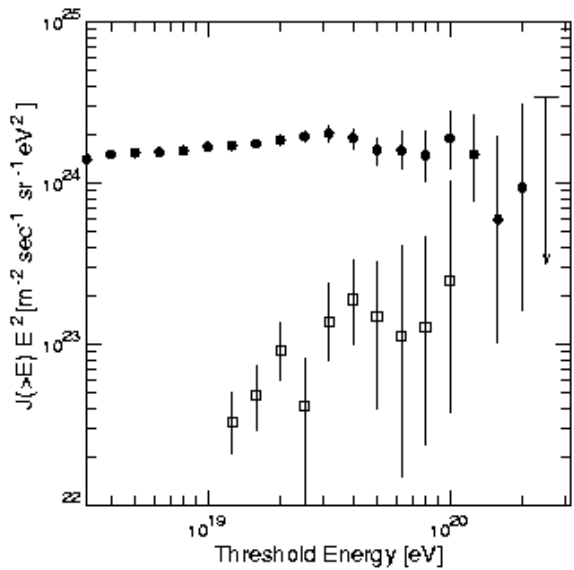}}   
\caption[...]{
AGASA integrated energy spectrum of cosmic rays (closed circles) and
their contribution to clustered events (open squares, see also Fig.~\ref{arr-dist}). 
From Ref.\cite{AGASA-clust}. 
\label{spect-clust}}
\end{figure}

The arrival directions of the cosmogenic neutrinos should pretty much coincide with the 
direction of the primary protons. Therefore, 
strongly interacting neutrino scenarios open a window of opportunity for the search for 
astrophysical point sources 
of post-GZK UHECR's located at cosmological distances\footnote{This window of opportunity
is shared with any scenario which exploits neutrinos as ``messenger'' particles, in particular also the Z-burst scenario.}.
In this context, it is interesting to note that AGASA observed a clustering of 
events on small angular scales\cite{Takeda:1999sg,AGASA-clust,AGASA-clust2} (cf. Fig.~\ref{arr-dist}) -- whose 
statistical significance of occurring higher than chance coincidence\cite{Goldberg:2000zq} is still being  
debated\cite{Bahcall:1999ap,Fodor:2000yi,Dubovsky:2000gv,HiRes-clust}, however. 
Intriguingly, the integrated flux of cosmic rays contributing to the AGASA event clusters, as shown 
by open squares in Fig.~\ref{spect-clust}, has a spectrum which is strikingly similar to the 
one expected from cosmogenic neutrinos in a strongly interacting neutrino scenario 
(cf. Fig.~\ref{fit} (dashed)).       
Possible correlations of the arrival distributions of UHECR's with definite distant astrophysical
sources such as compact radio quasars\cite{Farrar:1998we}, in particular BL Lacertae objects\cite{Tinyakov:2001nr}, 
or GRB's and magnetars\cite{Singh:2003xr}  
may give further circumstantial evidence for an UHEC$\nu$ component 
in EAS data. 
High statistics data from forthcoming cosmic ray facilities such as Auger\cite{Auger} and 
EUSO\cite{EUSO} are required\cite{Evans:2002ry} for these investigations.

\subsubsection{Searches at neutrino telescopes}

\begin{figure}[t]
\centerline{\epsfxsize=1.7in\epsfbox{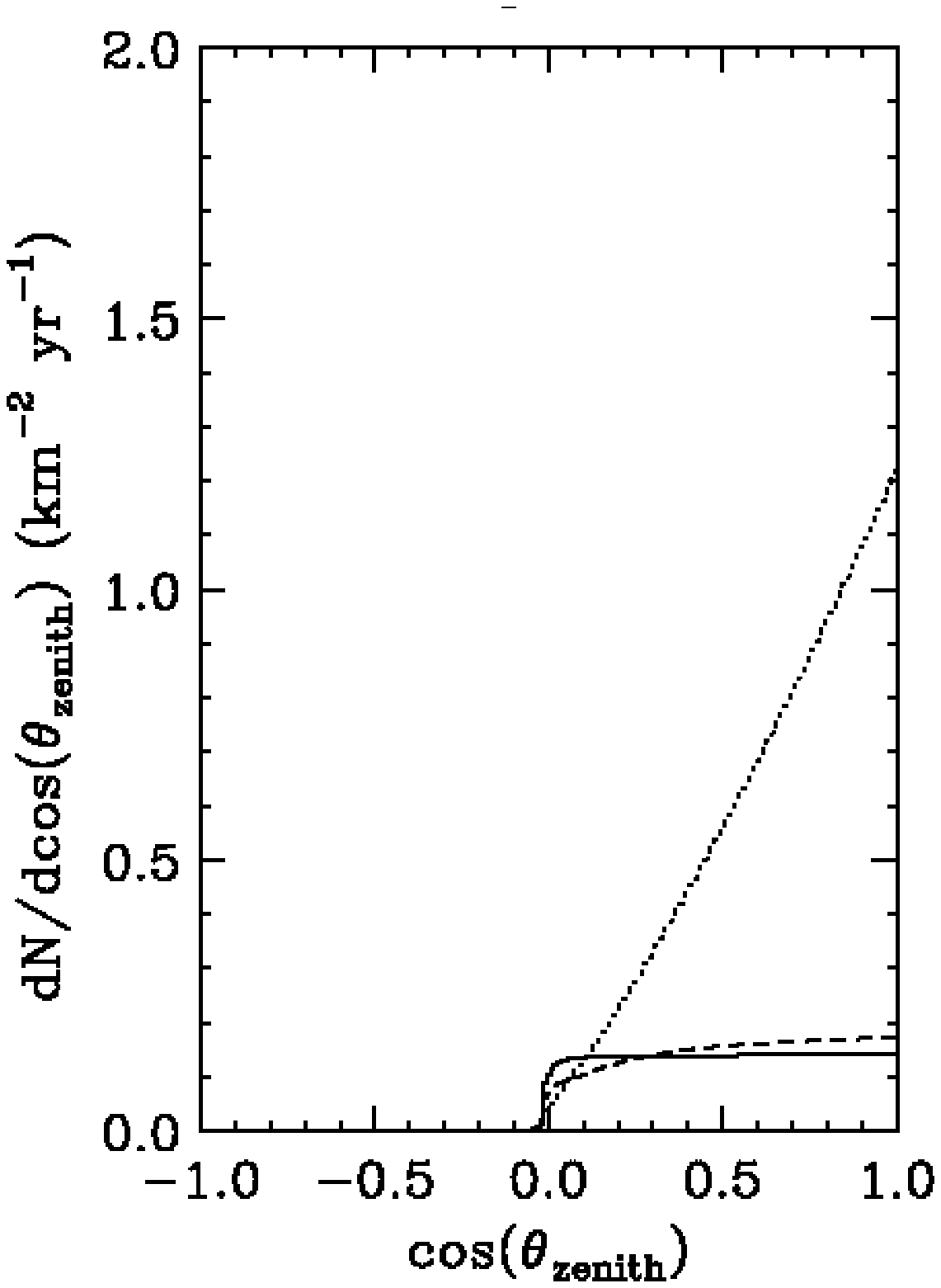}\hspace{8ex}\epsfxsize=1.9in\epsfbox{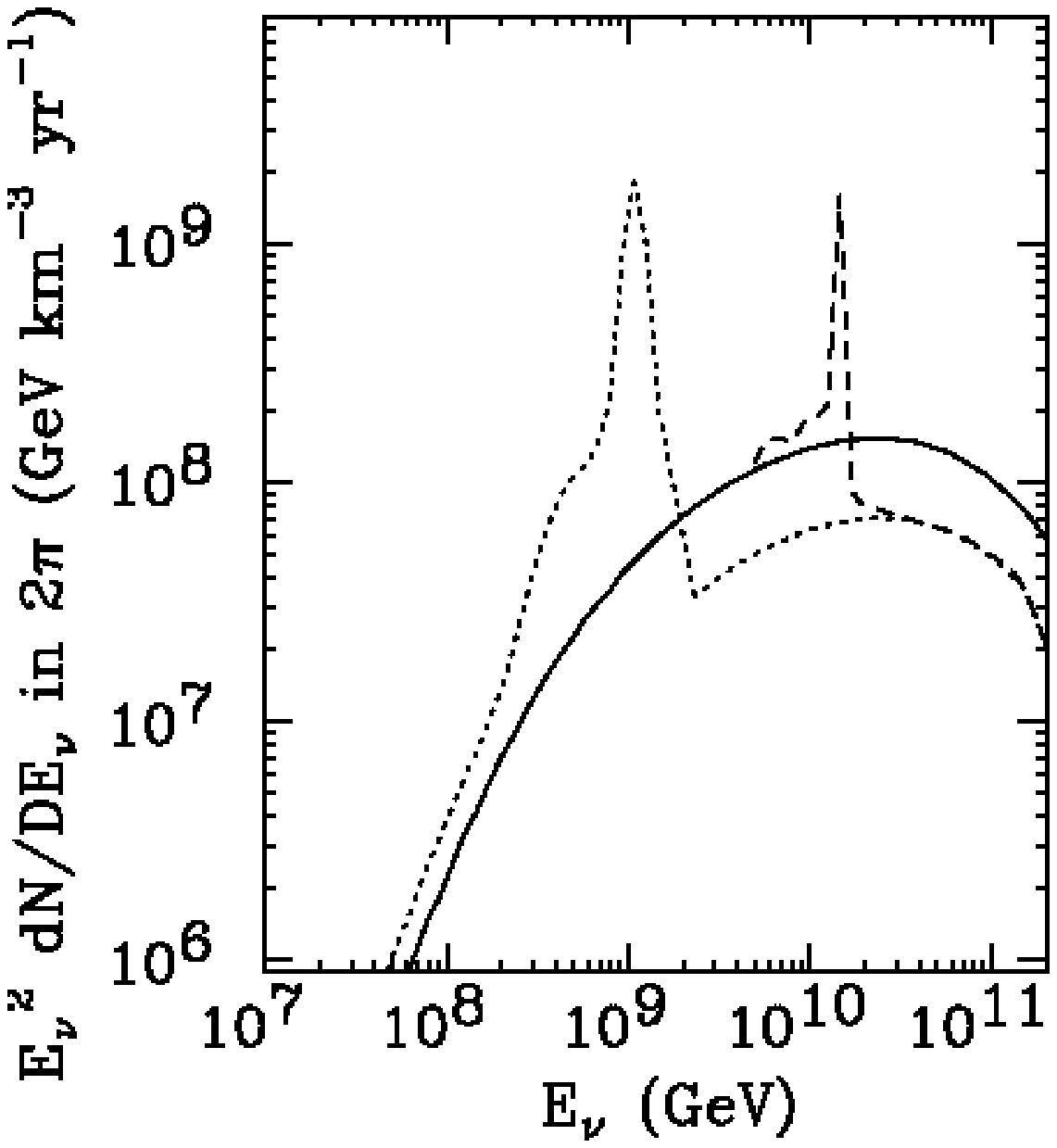}}   
\caption[...]{
Characteristic observables to be studied, in the context of 
strongly interacting neutrino scenarios, at a kilometer-scale neutrino detector\cite{Han:2003ru}
such as IceCube\cite{IceCube}.  
The cosmogenic neutrino flux from Ref.\cite{Engel:2001hd} 
was assumed. 
The dotted lines exploit the cross-section estimate from electroweak instantons 
from Refs.\cite{Ringwald:2002sw,Fodor:2003bn} (cf. Fig.~\ref{cross-nuN} (solid)), 
whereas the dashed lines represent the corresponding estimate from Refs.\cite{Bezrukov:2003er,Ringwald:2003ns} 
(cf. Fig.~\ref{cross-nuN} (dashed)). The solid lines are the Standard Model 
neutral+charged current predictions.
\\  
{\it Left:} 
Zenith angle distribution of showers generated in neutrino-nucleon 
interactions. 
A 1 EeV energy threshold for the observed showers has been imposed.\\  
{\it Right:}
Energy distribution of down-going showers generated in neutrino-nucleon 
interactions. 
\label{nt-obs}}
\end{figure}

The characteristic zenith angle distribution of showers in strongly interacting neutrino scenarios 
can of course be searched for also at neutrino telescopes where the absorbing material is, 
in addition to the atmosphere, the Earth as well as antartic ice (for IceCube\cite{IceCube}) or water 
(for ANTARES\cite{ANTARES}).  
This is illustrated in Fig.~\ref{nt-obs} (left), which displays the expected zenith angle distribution of 
neutrino-initiated showers above 1 EeV in a kilometer-scale detector\cite{Han:2003ru}.   
For Standard Model interactions, the distribution  (solid curve) 
is nearly flat for down-going events, and essentially no up-going events occur 
due to very efficient neutrino absorption by the Earth at these energies. 
For models with larger cross-sections, 
vertical down-going events become more frequent, producing
more events near $\cos\theta_{\rm zenith}\sim 1$. 
At zenith angles near the horizon, $\cos\theta_{\rm zenith}\sim 0$, 
more of the neutrinos are absorbed and the rate can be suppressed.

Another distinctive observable at neutrino telescopes is the energy spectrum of 
down-going shower events, which is shown in Fig.~\ref{nt-obs} (right) for the IceCube 
detector\cite{Han:2003ru}. The structure of this spectrum can be easily understood.  
As the neutrino energy exceeds the assumed threshold energy of the new interaction (cf. Fig.~\ref{cross-nuN}),
the number of events increases  dramatically above the Standard model prediction. 
Even farther above this energy, however, 
more of the neutrinos are absorbed in the ice
before reaching the detector and the event 
rate is suppressed.  This drastic ``bump" structure in the spectrum 
indicates the sharply enhanced cross-section at the  
threshold. The peak of this 
bump occurs at the associated neutrino threshold energy and is mainly generated 
by charged current electron neutrino interactions. The ``shoulder'' 
slightly to the left of the bump  is from neutral and charged 
current interactions which generate showers less energetic than the incident neutrino.

Let us mention also the possibility to look for enhanced rates for throughgoing 
muons (see Ref.\cite{Kowalski:2002gb} for a related study)  or even spatially compact 
muon bundles\cite{Morris:1991bb}. These signatures, however, rely on details of the final state from the new interaction
and are, therefore, more model-dependent than the ones discussed above, which exploit just
generic shower properties. 

\subsection{Laboratory tests}

\subsubsection{(Quasi-)elastic neutrino-(electron-)nucleon scattering}

As a consequence of dispersion relations, 
the hypothesized rapid rise in the neutrino-nucleon cross-section at large energies
is felt in the elastic neutrino-nucleon scattering amplitude at much lower 
energies\cite{Goldberg:1998pv}\footnote{This has been also pointed out in Ref.\cite{Burdman:1997yg}, however
in the context of perturbative and model-dependent considerations.}. 
Exploiting unitarity and analyticity, one may relate the invariant elastic $\nu N$ amplitudes 
$A_\pm (E)$, labeled by the nucleon helicity, with the total $\nu N$ cross-section via 
the dispersion relation  
\begin{equation} 
{\rm Re}\, A_\pm (E)-{\rm Re}\,A_\pm(0)
=\frac{E}{4\pi}\,\mathcal{P}
\int_0^{\infty}\ {\rm d}E^{\, \prime}\ \left(
\frac{\sigma_{\nu N}^{\rm tot}(E^{\, \prime},\pm)}{E^{\, \prime}(E^{\, \prime}-E)}\ +\
\frac{\sigma_{\bar \nu N}^{\rm tot}(E^{\, \prime},\pm)}{E^{\, \prime}(E^{\, \prime}+E)}
\right) \,,
\label{dispsubt} 
\end{equation}
where $\mathcal{P}$ denotes the principle
value of the integral. Suppose the new physics dominates the
neutrino-nucleon dispersion integral (\ref{dispsubt}) for
$E^{\, \prime}\ge E_{\rm th}$ as hypothesized in a strongly interacting neutrino scenario.
Assuming that
$\sigma^{\rm new}_{\nu N}$ is independent of helicity and energy, 
and obeys the Pomeranchuk relation  
$\sigma_{\nu N}^{\rm tot}(E,\pm)-\sigma_{\bar \nu
N}^{\rm tot}(E,\pm) \stackrel{E\rightarrow\infty}{\longrightarrow} 0$,
a new contribution to the real part of the amplitude at energy $E$ emerges from~(\ref{dispsubt}),
\begin{equation}
{\rm Re}\, A_\pm(E)\simeq \underbrace{{\rm Re}\, A_\pm(0)}_{G_F/2\sqrt{2}}
+ \frac{1}{2\pi}\frac{E}{E_{\rm th}}\sigma^{\rm new}_{\nu N}
\, ,  
\label{delta}
\end{equation}
resulting in 
\begin{equation}
\frac{{\rm Re}\,A(E)_{\rm new}}{{\rm Re}A(E)_{\rm SM}}
\simeq\left(\frac{E/100\ {\rm GeV}}{E_{\rm th}/10^{18}\ \mbox{eV}}\right)\
\left(\frac{\sigma^{\rm new}_{\nu N}}{100\ \rm{mb}}\right) 
\label{r}
\end{equation}
for the ratio of the new amplitude to the (perturbative) Standard Model (SM) amplitude. 
Order 100\,\% effects in
the real elastic amplitudes begin to appear already at energies seven orders of
magnitude below the full realization of the strong cross-section. 

Neutrino-nucleon scattering experiments in the laboratory have therefore the 
opportunity to test strongly interacting neutrino scenarios by searching
for enhancements in the elastic cross-sections. Current experiments
at CERN and Fermilab reach energies around 100 GeV and, therefore, already 
start to constrain possible scenarios, e.g. $p$-brane production\cite{Anchordoqui:2002it}.   
Elastic and quasi-elastic 
scattering processes may be studied with the help of  
the H1\cite{H1} and ZEUS\cite{ZEUS} detectors at the HERA $e^\pm p$ collider.
Its $e^\pm$ energy, in the proton's restframe, is around $10^5$~GeV and, therefore, 
extends much beyond the energy reach of the above mentioned $\nu N$ fixed-target 
experiments. However, one-photon exchange dominates the low-energy 
elastic amplitude for $e^\pm p\rightarrow e^\pm p$ to such an extent that the anomalous, new contribution is 
suppressed by a factor of $\sim 1/100$ compared to (\ref{r}).   
On the other hand, possible enhancements in the quasi-elastic channels $e^+ p\rightarrow \nu_e\, n$ and 
$e^- p\rightarrow \bar{\nu}_e\, n$, which do not suffer from QED dominance, cannot 
be deduced from model-independent dispersion relations. A separate calculation could be made, however, 
if certain aspects of the new high-energy strong-interaction are assumed.   
 
\subsubsection{Instanton searches}

There is a close analogy\cite{Balitsky:1993jd} between electroweak and hard QCD 
instanton-induced processes in deep-inelastic scattering\cite{Moch:1996bs}. 
An observably large cross-section for the latter processes at HERA is indeed 
necessary, but not sufficient for an observably large cross-section for the 
former. It seems, moreover, that a $\gwig\, {\rm mb}$ electroweak instanton cross-section 
necessarily requires that the bulk of inelastic small-Bjorken-$x$ processes is induced by 
soft QCD instantons, as has been proposed indeed in Ref.\cite{Kharzeev:2000ef}. 
Present upper limits on hard QCD instantons from 
the H1\cite{Adloff:2002ph} and ZEUS\cite{Chekanov:2003ww} experiments are still above the theoretical 
predictions, but may be improved considerably at HERA II, within this decade.    
The possible direct production and observation\cite{Farrar:1990vb} of electroweak instanton induced processes 
in the laboratory 
will have to wait for the commissioning of the VLHC\cite{Ringwald:2002sw}.

\section{\label{conclusions}Summary and conclusions}

We have shown that a simple scenario with a single power-law  
injection spectrum of protons can describe all 
the features of the UHECR spectrum 
in the energy range $10^{17\div 21}$~eV, provided the neutrino-nucleon
cross-sections becomes of hadronic size at energies above $\approx 10^{19}$~eV.  
In such a strongly interacting neutrino scenario, the 
cosmogenic neutrinos, which have been produced during proton propagation through
the CMB, initiate air showers high up in the atmosphere and give thus rise
to a second, neutrino-induced EAS component, extending well beyond the GZK energy. 
As examples giving rise to the necessary enhancement in $\sigma_{\nu N}$, we discussed 
$p$-brane production in TeV-scale gravity scenarios and Standard Model electroweak instanton-induced
processes. 
The model for the proton injection spectrum has few parameters from which only two -- 
the power index $\alpha$ and the redshift evolution index $n$ -- 
has a strong effect on the final shape of the spectrum.
We found that, for certain values of $\alpha$ and $n$, strongly interacting neutrino scenarios are 
compatible with the available observational data from the AGASA and HiRes experiments 
(combined with their predecessor  experiments, 
Fly's Eye and Akeno, respectively) on the 2-sigma level (also 1-sigma for
HiRes). 
There are a number of astrophysical source candidates, notably  neutron stars 
and GRB's, which may provide the necessary conditions to 
accelerate protons to the required energies, $E_{\rm max}\gwig\, 3\times 10^{21}$~eV, 
by conventional shock acceleration.    

The predicted ultrahigh energy cosmic neutrino component can be experimentally tested by 
studying the zenith angle dependence of the events in the range 
$10^{18\div 20}$~eV and possible correlations with distant
astrophysical sources at cosmic ray facilities such as the Pierre Auger Observatory and EUSO, 
and by looking for bumps in neutrino-initiated shower spectra 
at neutrino telescopes such as ANTARES and IceCube.  
As laboratory tests, one may search for a enhancements  
in (quasi-)elastic lepton-nucleon scattering 
or for signatures of QCD instanton-induced processes in deep-inelastic
scattering, e.g. at HERA. 

In summary, strongly interacting neutrino scenarios provide a viable 
and attractive solution to the ultrahigh energy cosmic ray puzzle and
may be subject to various crucial tests in the foreseeable future. 

\section*{Acknowledgments}
We would like to thank Luis Anchordoqui for valuable comments and discussions.


\begin{thebibliography}{0}

\bibitem{Olinto:2000fz}
A.~V.~Olinto,
in: Proc. International Workshop on {\it Observing Ultra High Energy Cosmic Rays from Space and on Earth}, 
Metepec, Puebla, Mexico, 9-12 Aug 2000, published in: AIP Conf.Proc.566:99-112,2000  
[arXiv:astro-ph/0102077].

\bibitem{Cronin:vy}
J.~W.~Cronin, S.~P.~Swordy and T.~K.~Gaisser,
{\it Sci.\ Am.}\  {\bf 276}, 32 (1997).

\bibitem{vlhc}
Very Large Hadron Collider, 
http://vlhc.org . 

\bibitem{lhc}
Large Hadron Collider, 
http://lhc-new-homepage.web.cern.ch/lhc-new-homepage . 

\bibitem{Anchordoqui:2002hs}
L.~Anchordoqui, T.~Paul, S.~Reucroft and J.~Swain,
{\it Int.\ J.\ Mod.\ Phys.}\ A {\bf 18}, 2229 (2003).

\bibitem{AGASA}
Akeno Giant Air Shower Array, 
http://www-akeno.icrr.u-tokyo.ac.jp/AGASA .

\bibitem{Takeda:1999sg}
M.~Takeda {\it et al.},
{\it Astrophys.\ J.}\  {\bf 522}, 225 (1999).

\bibitem{Greisen:1966jv}
K.~Greisen,
{\it Phys.\ Rev.\ Lett.}\  {\bf 16}, 748 (1966); 
%
G.~T.~Zatsepin and V.~A.~Kuzmin,
{\it JETP Lett.}\  {\bf 4}, 78 (1966) 
[{\it Pisma Zh.\ Eksp.\ Teor.\ Fiz.}\  {\bf 4}, 114 (1966)].

\bibitem{Nagano:1991jz}
M.~Nagano {\it et al.},
{\it J.\ Phys.}\ G {\bf 18}, 423 (1992).
%

\bibitem{Takeda:1998ps}
M.~Takeda {\it et al.},
{\it Phys.\ Rev.\ Lett.}\  {\bf 81}, 1163 (1998); 
http://www-akeno.icrr.u-tokyo.ac.jp/AGASA ; 
date: February 24, 2003.

\bibitem{Bird:yi}
D.~J.~Bird {\it et al.},  
{\it Phys.\ Rev.\ Lett.}\  {\bf 71}, 3401 (1993); 
%
D.~J.~Bird {\it et al.}  [HIRES Collaboration],
{\it Astrophys.\ J.}\  {\bf 424}, 491 (1994); 
%
D.~J.~Bird {\it et al.},
{\it Astrophys.\ J.}\  {\bf 441}, 144 (1995).


\bibitem{Abu-Zayyad:2002ta}
T.~Abu-Zayyad {\it et al.}  
[HiRes Collaboration], 
astro-ph/0208243; 
%
astro-ph/0208301.


\bibitem{Fodor:2003ph}
Z.~Fodor, S.~D.~Katz, A.~Ringwald and H.~Tu,
{\it JCAP} {\bf 0311}, 015 (2003).

\bibitem{Ginzburg:63} 
V.~L.~Ginzburg and S.~I.~Syrovatsky,
{\it The origin of cosmic rays}, (Pergamon Press, Oxford 1964);
%
J.~Wdowczyk and A.~W.~Wolfendale,
{\it Nature} {\bf 281}, 356 (1979).

\bibitem{Ahn:1999jd}
E.~J.~Ahn, G.~A.~Medina-Tanco, P.~L.~Biermann and T.~Stanev,
astro-ph/9911123;
%
P.~L.~Biermann, E.~J.~Ahn, G.~A.~Medina-Tanco and T.~Stanev,
{\it Nucl.\ Phys.\ Proc.\ Suppl.}\  {\bf 87}, 417 (2000).

\bibitem{Billoir:2000wi}
P.~Billoir and A.~Letessier-Selvon,
astro-ph/0001427.

\bibitem{Biermann} P. L. Biermann, E. -J. Ahn, P. P. Kronberg, 
G. Medina Tanco, and T. Stanev,
in: {\it Physics and Astrophysics of Ultra-High-Energy Cosmic Rays},
edts. M. Lemoine and G. Sigl, Springer-Verlag, Berlin, 2001.

\bibitem{Beresinsky:qj}
V.~S.~Beresinsky and G.~T.~Zatsepin,
{\it Phys.\ Lett.}\ B {\bf 28}, 423 (1969); 
%
{\it Sov.\ J.\ Nucl.\ Phys.}\  {\bf 11}, 111 (1970)
[{\it Yad.\ Fiz.}\ {\bf 11}, 200 (1970)].


\bibitem{Weiler:1997sh}
T.~J.~Weiler,
{\it Astropart.\ Phys.}\  {\bf 11}, 303 (1999); 
%
D.~Fargion {\em et al.},
{\it Astrophys.\ J.}\  {\bf 517}, 725 (1999);
%
S.~Yoshida, G.~Sigl and S.~J.~Lee,
{\it Phys.\ Rev.\ Lett.}\  {\bf 81}, 5505 (1998). 


\bibitem{Fodor:2001qy}
Z.~Fodor, S.~D.~Katz and A.~Ringwald,
{\it Phys.\ Rev.\ Lett.}\  {\bf 88}, 171101 (2002);   
%
{\it JHEP} {\bf 0206}, 046 (2002); 
%
O.~E.~Kalashev, V.~A.~Kuzmin, D.~V.~Semikoz and G.~Sigl,
{\it Phys.\ Rev.}\ D {\bf 65}, 103003 (2002); 
%
D.~V.~Semikoz and G.~Sigl,
hep-ph/0309328. 



\bibitem{Kravchenko:2003tc}
I.~Kravchenko,
astro-ph/0306408;
%
P.~W.~Gorham, C.~L.~Hebert, K.~M.~Liewer, C.~J.~Naudet, D.~Saltzberg and D.~Williams,
astro-ph/0310232; 
%
N.~G.~Lehtinen, P.~W.~Gorham, A.~R.~Jacobson and R.~A.~Roussel-Dupre,
astro-ph/0309656.


\bibitem{Auger}
Pierre Auger Observatory, 
http://www.auger.org . 

\bibitem{IceCube}
IceCube, 
http://icecube.wisc.edu . 


\bibitem{ANITA}
ANtarctic Impulse Transient Array,   
http://www.ps.uci.edu/\~{}anita . 

\bibitem{EUSO}
Extreme Universe Space Observatory,  
http://www.euso-mission.org .

\bibitem{OWL}
Orbiting Wide-angle Light-collectors,  
http://owl.gsfc.nasa.gov .

\bibitem{Gorham:2001wr}
Saltdome Shower Array, 
P.~Gorham
{\em \ et al.}, 
{\it Nucl.\ Instrum.\ Meth.}\ A {\bf 490}, 476 (2002).

\bibitem{Eberle:2004ua}
B.~Eberle, A.~Ringwald, L.~Song and T.~J.~Weiler,
hep-ph/0401203.



\bibitem{Stecker:1979ah}
F.~W.~Stecker,
{\it Astrophys.\ J.}\  {\bf 228}, 919 (1979);
%
C.~T.~Hill and D.~N.~Schramm,
{\it Phys.\ Rev.}\ D {\bf 31}, 564 (1985);
%
C.~T.~Hill, D.~N.~Schramm and T.~P.~Walker,
{\it Phys.\ Rev.}\ D {\bf 34}, 1622 (1986);
%
F.~W.~Stecker, C.~Done, M.~H.~Salamon and P.~Sommers,
{\it Phys.\ Rev.\ Lett.}\  {\bf 66}, 2697 (1991)
[Erratum-ibid.\  {\bf 69}, 2738 (1991)].

\bibitem{Yoshida:pt}
S.~Yoshida and M.~Teshima,
{\it Prog.\ Theor.\ Phys.}\  {\bf 89}, 833 (1993).


\bibitem{Protheroe:1995ft}
R.~J.~Protheroe and P.~A.~Johnson,
{\it Astropart.\ Phys.}\  {\bf 4}, 253 (1996); 
%
S.~Yoshida, H.~Y.~Dai, C.~C.~H.~Jui and P.~Sommers,
{\it Astrophys.\ J.}\  {\bf 479}, 547 (1997);
%
R.~Engel and T.~Stanev,
{\it Phys.\ Rev.}\ D {\bf 64} (2001) 093010; 
%
O.~E.~Kalashev, V.~A.~Kuzmin, D.~V.~Semikoz and G.~Sigl,
{\it Phys.\ Rev.}\ D {\bf 66}, 063004 (2002); 
%
Z.~Fodor, S.~D.~Katz, A.~Ringwald and H.~Tu,
{\it JCAP} {\bf 0311}, 015 (2003);
%
D.~Semikoz and G.~Sigl,
hep-ph/0309328.

\bibitem{Bordes:1997bt}
J.~Bordes, H.~M.~Chan, J.~Faridani, J.~Pfaudler and S.~T.~Tsou,
hep-ph/9705463;
%
{\it Astropart.\ Phys.}\  {\bf 8}, 135 (1998).


\bibitem{Domokos:2000dp}
G.~Domokos, S.~Kovesi-Domokos and P.~T.~Mikulski,
hep-ph/0006328.

\bibitem{Domokos:1998ry}
G.~Domokos and S.~Kovesi-Domokos,
{\it Phys.\ Rev.\ Lett.}\  {\bf 82}, 1366 (1999);
%
S.~Nussinov and R.~Shrock,
{\it Phys.\ Rev.}\ D {\bf 59}, 105002 (1999); 
%
P.~Jain, D.~W.~McKay, S.~Panda and J.~P.~Ralston,
{\it Phys.\ Lett.}\ B {\bf 484}, 267 (2000); 
%
A.~V.~Kisselev and V.~A.~Petrov,
hep-ph/0311356.

\bibitem{Kachelriess:2000cb}
M.~Kachelriess and M.~Pl\"umacher,
{\it Phys.\ Rev.}\ D {\bf 62}, 103006 (2000); 
%
L.~Anchordoqui, H.~Goldberg, T.~McCauley, T.~Paul, S.~Reucroft and J.~Swain,
{\it Phys.\ Rev.}\ D {\bf 63}, 124009 (2001).

\bibitem{Ahn:2002mj}
E.~J.~Ahn, M.~Cavaglia and A.~V.~Olinto,
{\it Phys.\ Lett.}\ B {\bf 551}, 1 (2003);
%
P.~Jain, S.~Kar, S.~Panda and J.~P.~Ralston,
{\it Int.\ J.\ Mod.\ Phys.}\ D {\bf 12}, 1593 (2003).

\bibitem{Anchordoqui:2002it}
L.~A.~Anchordoqui, J.~L.~Feng and H.~Goldberg,
{\it Phys.\ Lett.}\ B {\bf 535}, 302 (2002).


\bibitem{Domokos:1986qy}
G.~Domokos and S.~Nussinov,
{\it Phys.\ Lett.}\ B {\bf 187}, 372 (1987).



\bibitem{Barshay:2001eq}
S.~Barshay and G.~Kreyerhoff,
{\it Eur.\ Phys.\ J.}\ C {\bf 23}, 191 (2002);
%
{\it Phys.\ Lett.}\ B {\bf 535}, 201 (2002);
%
S.~Kovesi-Domokos and G.~Domokos,
hep-ph/0307098; 
%
hep-ph/0307099.


\bibitem{Fodor:2003bn}
Z.~Fodor, S.~D.~Katz, A.~Ringwald and H.~Tu,
{\it Phys.\ Lett.}\ B {\bf 561}, 191 (2003).

\bibitem{Aoyama:1986ej}
H.~Aoyama and H.~Goldberg,
{\it Phys.\ Lett.}\ B {\bf 188}, 506 (1987);
%
A.~Ringwald,
{\it Nucl.\ Phys.}\ B {\bf 330}, 1 (1990);  
%
O.~Espinosa,
{\it Nucl.\ Phys.}\ B {\bf 343}, 310 (1990);  
%
V.~V.~Khoze and A.~Ringwald,
{\it Phys.\ Lett.}\ B {\bf 259}, 106 (1991). 
%

\bibitem{Morris:1991bb}
D.~A.~Morris and R.~Rosenfeld,
{\it Phys.\ Rev.}\ D {\bf 44}, 3530 (1991);
%
D.~A.~Morris and A.~Ringwald,
{\it Astropart.\ Phys.}\  {\bf 2}, 43 (1994).


\bibitem{Ringwald:2002sw}
A.~Ringwald,
{\it Phys.\ Lett.}\ B {\bf 555}, 227 (2003).



\bibitem{Bahcall:1999ap}
J.~N.~Bahcall and E.~Waxman,
{\it Astrophys.\ J.}\  {\bf 542}, 543 (2000).

\bibitem{Fodor:2000yi}
Z.~Fodor and S.~D.~Katz,
{\it Phys.\ Rev.}\ D {\bf 63}, 023002 (2001).

\bibitem{Waxman:1995dg}
E.~Waxman,
{\it Astrophys.\ J.}\  {\bf 452}, L1 (1995).

\bibitem{Hagiwara:fs}
K.~Hagiwara {\it et al.}  [Particle Data Group],
{\it Phys.\ Rev.}\ D {\bf 66}, 010001 (2002).

\bibitem{Mucke:1999yb}
A.~M\"ucke, R.~Engel, J.~P.~Rachen, R.~J.~Protheroe and T.~Stanev,
{\it Comput.\ Phys.\ Commun.}\  {\bf 124}, 290 (2000).

\bibitem{P:xxx}
Z.~Fodor, S.~D.~Katz and A.~Ringwald, in preparation.

\bibitem{Bertou:2001vm}
X.~Bertou, P.~Billoir, O.~Deligny, C.~Lachaud and A.~Letessier-Selvon,
{\it Astropart.\ Phys.}\  {\bf 17}, 183 (2002);
%
C.~Lachaud, X.~Bertou, P.~Billoir, O.~Deligny and A.~Letessier-Selvon,
{\it Nucl.\ Phys.\ Proc.\ Suppl.}\  {\bf 110}, 525 (2002).

\bibitem{Antoniadis:1990ew}
I.~Antoniadis,
{\it Phys.\ Lett.}\ B {\bf 246}, 377 (1990); 
J.~D.~Lykken,
{\it Phys.\ Rev.}\ D {\bf 54}, 3693 (1996); 
N.~Arkani-Hamed, S.~Dimopoulos and G.~R.~Dvali,
{\it Phys.\ Lett.}\ B {\bf 429}, 263 (1998); 
L.~Randall and R.~Sundrum,
{\it Phys.\ Rev.\ Lett.}\  {\bf 83}, 3370 (1999). 

\bibitem{Giddings:2001bu}
S.~B.~Giddings and S.~Thomas,
{\it Phys.\ Rev.}\ D {\bf 65}, 056010 (2002);
%
S.~Dimopoulos and G.~Landsberg,
{\it Phys.\ Rev.\ Lett.}\  {\bf 87}, 161602 (2001).

\bibitem{Feng:2001ib}
J.~L.~Feng and A.~D.~Shapere,
{\it Phys.\ Rev.\ Lett.}\  {\bf 88}, 021303 (2002);
%
A.~Ringwald and H.~Tu,
{\it Phys.\ Lett.}\ B {\bf 525}, 135 (2002); 
%
L.~A.~Anchordoqui {\it et al.}, 
hep-ph/0309082.

\bibitem{Han:2003ru}
T.~Han and D.~Hooper,
{\it Phys.\ Lett.}\ B {\bf 582}, 21 (2004).

\bibitem{Gandhi:1998ri}
R.~Gandhi, C.~Quigg, M.~H.~Reno and I.~Sarcevic,
{\it Phys.\ Rev.}\ D {\bf 58}, 093009 (1998).

\bibitem{Bezrukov:2003er}
F.~Bezrukov {\em et al.},
{\it Phys.\ Rev.}\ D {\bf 68}, 036005 (2003). 

\bibitem{Ringwald:2003ns}
A.~Ringwald,
{\it JHEP} {\bf 0310}, 008 (2003).


\bibitem{Kachelriess:2003yy}
M.~Kachelriess, D.~V.~Semikoz and M.~A.~Tortola,
{\it Phys.\ Rev.}\ D {\bf 68}, 043005 (2003).

\bibitem{Bahcall:2002wi}
J.~N.~Bahcall and E.~Waxman,
{\it Phys.\ Lett.}\ B {\bf 556}, 1 (2003).


\bibitem{DeMarco:2003ig}
D.~De Marco, P.~Blasi and A.~V.~Olinto,
{\it Astropart.\ Phys.}\  {\bf 20}, 53 (2003).

\bibitem{Hillas:1984} 
A.~M.~Hillas,
Ann. Rev. Astron. Astrophys. {\bf 22}, 425 (1984).


\bibitem{Berezinsky:kz}
V.~S.~Berezinsky and A.~Y.~Smirnov,
{\it Phys.\ Lett.}\ B {\bf 48}, 269 (1974);
%
C.~Tyler, A.~V.~Olinto and G.~Sigl,
{\it Phys.\ Rev.}\ D {\bf 63}, 055001 (2001);
%
A.~Kusenko and T.~J.~Weiler,
{\it Phys.\ Rev.\ Lett.}\  {\bf 88}, 161101 (2002).


\bibitem{Baltrusaitis:mt}
R.~M.~Baltrusaitis {\it et al.},
{\it Phys.\ Rev.}\ D {\bf 31}, 2192 (1985).


\bibitem{Yoshida:2001icrc}
S.~Yoshida {\it et al.} [AGASA Collaboration], in:
{\it Proc. 27th International Cosmic Ray Conference}, 
Hamburg, Germany, 2001, Vol. 3, p. 1142.

\bibitem{AGASA-clust}
M.~Takeda {\it et al.} [AGASA Collaboration], 
in:
{\it Proc. 27th International Cosmic Ray Conference}, 
Hamburg, Germany, 2001, Vol. 3, p. 345.

\bibitem{AGASA-clust2}
M.~Teshima {\it et al.} [AGASA Collaboration], 
in:
{\it Proc. 28th International Cosmic Ray Conference}, 
Tsukuba, Japan, 2003, p. 437.

\bibitem{Goldberg:2000zq}
H.~Goldberg and T.~J.~Weiler,
{\it Phys.\ Rev.}\ D {\bf 64}, 056008 (2001).

\bibitem{Dubovsky:2000gv}
S.~L.~Dubovsky, P.~G.~Tinyakov and I.~I.~Tkachev,
{\it Phys.\ Rev.\ Lett.}\  {\bf 85}, 1154 (2000);
%
P.~G.~Tinyakov and I.~I.~Tkachev,
{\it JETP Lett.}\  {\bf 74}, 1 (2001)
[{\it Pisma Zh.\ Eksp.\ Teor.\ Fiz.}\  {\bf 74}, 3 (2001)];
%
Y.~Uchihori, M.~Nagano, M.~Takeda, M.~Teshima, J.~Lloyd-Evans and A.~A.~Watson,
{\it Astropart.\ Phys.}\  {\bf 13}, 151 (2000);
%
L.~A.~Anchordoqui, H.~Goldberg, S.~Reucroft, G.~E.~Romero, J.~Swain and D.~F.~Torres,
{\it Mod.\ Phys.\ Lett.}\ A {\bf 16}, 2033 (2001);
%
C.~B.~Finley and S.~Westerhoff,
astro-ph/0309159.

\bibitem{HiRes-clust}
C.~Finley {\it et al.} [HiRes Collaboration], 
in:
{\it Proc. 28th International Cosmic Ray Conference}, 
Tsukuba, Japan, 2003, p. 433.

\bibitem{Farrar:1998we}
G.~R.~Farrar and P.~L.~Biermann,
{\it Phys.\ Rev.\ Lett.}\  {\bf 81}, 3579 (1998).

\bibitem{Tinyakov:2001nr}
P.~G.~Tinyakov and I.~I.~Tkachev,
{\it JETP Lett.}\  {\bf 74}, 445 (2001)
[{\it Pisma Zh.\ Eksp.\ Teor.\ Fiz.}\  {\bf 74}, 499 (2001)]; 
%
{\it Astropart.\ Phys.}\  {\bf 18}, 165 (2002);
%
D.~S.~Gorbunov, P.~G.~Tinyakov, I.~I.~Tkachev and S.~V.~Troitsky,
{\it Astrophys.\ J.}\  {\bf 577}, L93 (2002).

\bibitem{Singh:2003xr}
S.~Singh, C.~P.~Ma and J.~Arons,
astro-ph/0308257.

\bibitem{Evans:2002ry}
N.~W.~Evans, F.~Ferrer and S.~Sarkar,
{\it Phys.\ Rev.}\ D {\bf 67}, 103005 (2003);
%
D.~F.~Torres, S.~Reucroft, O.~Reimer and L.~A.~Anchordoqui,
{\it Astrophys.\ J.}\  {\bf 595}, L13 (2003);
%
J.~Swain,
astro-ph/0401632.

\bibitem{ANTARES}
Astronomy with a Neutrino Telescope and Abyss environmental RESearch,\\ 
http://antares.in2p3.fr .



\bibitem{Engel:2001hd}
R.~Engel, D.~Seckel and T.~Stanev,
{\it Phys.\ Rev.}\ D {\bf 64}, 093010 (2001).

\bibitem{Kowalski:2002gb}
M.~Kowalski, A.~Ringwald and H.~Tu,
{\it Phys.\ Lett.}\ B {\bf 529}, 1 (2002).

\bibitem{Goldberg:1998pv}
H.~Goldberg and T.~J.~Weiler,
{\it Phys.\ Rev.}\ D {\bf 59}, 113005 (1999).

\bibitem{Burdman:1997yg}
G.~Burdman, F.~Halzen and R.~Gandhi,
{\it Phys.\ Lett.}\ B {\bf 417}, 107 (1998).

\bibitem{H1}
H1 experiment at HERA, 
http://www-h1.desy.de .

\bibitem{ZEUS}
ZEUS experiment at HERA,
http://www-zeus.desy.de . 

\bibitem{Balitsky:1993jd}
I.~I.~Balitsky and V.~M.~Braun,
{\it Phys.\ Lett.}\ B {\bf 314}, 237 (1993); 
%
A.~Ringwald and F.~Schrempp,
in: {\em Quarks '94}, 
Vladimir, Russia, 1994, 
hep-ph/9411217.

\bibitem{Moch:1996bs}
S.~Moch, A.~Ringwald and F.~Schrempp,
{\it Nucl.\ Phys.}\ B {\bf 507}, 134 (1997); 
%
A.~Ringwald and F.~Schrempp,
{\it Phys.\ Lett.}\ B {\bf 438}, 217 (1998); 
%
{\it Comput.\ Phys.\ Commun.}\  {\bf 132}, 267 (2000); 
%
{\it Phys.\ Lett.}\ B {\bf 503}, 331 (2001).


\bibitem{Kharzeev:2000ef}
D.~E.~Kharzeev, Y.~V.~Kovchegov and E.~Levin,
{\it Nucl.\ Phys.}\ A {\bf 690}, 621 (2001); 
%
M.~A.~Nowak, E.~V.~Shuryak and I.~Zahed,
{\it Phys.\ Rev.}\ D {\bf 64}, 034008 (2001);   
%
F.~Schrempp,
{\it J.\ Phys.}\ G {\bf 28}, 915 (2002);
%
F.~Schrempp and A.~Utermann,
{\it Acta Phys.\ Polon.}\ B {\bf 33}, 3633 (2002); 
%
{\it Phys.\ Lett.}\ B {\bf 543}, 197 (2002); 
%
hep-ph/0301177;
%
hep-ph/0401137.

\bibitem{Adloff:2002ph}
C.~Adloff {\it et al.}  [H1 Collaboration],
{\it Eur.\ Phys.\ J.}\ C {\bf 25}, 495 (2002).

\bibitem{Chekanov:2003ww}
S.~Chekanov {\it et al.}  [ZEUS Collaboration],
hep-ex/0312048.

\bibitem{Farrar:1990vb}
G.~R.~Farrar and R.-B.~Meng,
{\it Phys.\ Rev.\ Lett.}\  {\bf 65}, 3377 (1990);
%
A.~Ringwald, F.~Schrempp and C.~Wetterich,
{\it Nucl.\ Phys.}\ B {\bf 365}, 3 (1991); 
%
M.~J.~Gibbs, A.~Ringwald, B.~R.~Webber and J.~T.~Zadrozny,
{\it Z.\ Phys.}\ C {\bf 66}, 285 (1995);
%
M.~J.~Gibbs and B.~R.~Webber,
{\it Comput.\ Phys.\ Commun.}\  {\bf 90}, 369 (1995).


\end{thebibliography}
\end{document}